\documentclass{sigplanconf}

\usepackage{amsmath}
\usepackage{color}
\usepackage{textcomp}
\usepackage{pgfplots}
\usepackage{pgfplotstable}
\usepackage{subcaption}
\usepackage{listings}
\usepackage{multirow}
\usepackage[T1]{fontenc}
\usepackage[scaled=0.81]{beramono} 
\usepackage{flushend}

\newcommand{\code}[1]{\lstinline[basicstyle=\small]{#1}}
\newcommand{\scode}[1]{\lstinline[basicstyle=\scriptsize]{#1}}

\newcommand*{\lstitem}[1]{
  \setbox0\hbox{\lstinline[basicstyle=\small]{#1}}  
  \item[\usebox0]  
  \hfill \\
}

\usepackage{graphicx} 
\usepackage{xcolor} 
\usepackage{xspace}
\usepackage{pgf,tikz}
\usetikzlibrary{calc,shapes,positioning,arrows,shapes.multipart,fit}

\newcommand{\thename}{SafeGPU\xspace}

\newcommand{\tabref}[1]{Table~\ref{tab:#1}}
\newcommand{\secref}[1]{Section~\ref{sec:#1}}
\newcommand{\lstref}[1]{Listing~\ref{#1}}

\usepackage{float}
\floatstyle{plain}
\newfloat{Listing}{htb}{prg}

\usepackage{url}

\begin{document}

\toappear{}

\setlength{\pdfpageheight}{\paperheight}
\setlength{\pdfpagewidth}{\paperwidth}

\preprintfooter{Preprint submitted to GPCE 2015}

\title{Contract-Based General-Purpose GPU Programming}

\authorinfo{Alexey Kolesnichenko\and Christopher M. Poskitt\and Sebastian Nanz\and Bertrand Meyer}{Department of Computer Science, ETH Z\"{u}rich, Switzerland}{firstname.lastname@inf.ethz.ch}

\maketitle

\begin{abstract}
Using GPUs as general-purpose processors has revolutionized parallel computing by offering, for a large and growing set of algorithms, massive data-parallelization on desktop machines. An obstacle to widespread adoption, however, is the difficulty of programming them and the low-level control of the hardware required to achieve good performance.
This paper suggests a programming library, \thename, that aims at striking a balance between programmer productivity and performance, by making GPU data-parallel operations accessible from within a classical object-oriented programming language. The solution is integrated with the design-by-contract approach, which increases confidence in functional program correctness by embedding executable program specifications into the program text.  
We show that our library leads to modular and maintainable code that is accessible to GPGPU non-experts, while providing performance that is comparable with hand-written CUDA code. Furthermore, runtime contract checking turns out to be feasible, as the contracts can be executed on the GPU.
\end{abstract}



\section{Introduction}
\label{sec:intro}

Graphics Processing Units (GPUs) are being increasingly leveraged as sources of inexpensive parallel-processing power, with application areas as diverse as scientific data analysis, cryptography, and evolutionary computing~\cite{NVIDIA-Applications,YooHU13a}. Consisting of thousands of processors, GPUs are throughput-oriented systems that are especially well-suited to realizing data-parallel algorithms---algorithms performing the same tasks on multiple items of data---with potentially significant performance gains to be achieved.

The CUDA~\cite{NVIDIA-CUDA} and OpenCL~\cite{OpenCL12a} languages support the programming of GPUs for applications beyond graphics in an approach now known as General-Purpose Computing on GPUs (GPGPU). They provide programmers with fine-grained control over hardware at the C++ level of abstraction. This control, however, is a double-edged sword: while it facilitates advanced, hardware-specific fine-tuning techniques, it does so at the cost of working within very restrictive and low-level programming models. Recursion, for example, is among the standard programming concepts prohibited. Furthermore, dynamic memory management is completely absent, meaning that programmers themselves must explicitly manage the allocation of memory and the movement of data. Although acceptable for specialist programmers, these issues pose a significant difficulty to others, and are an obstacle to more widespread adoption.

Such challenges have not gone unnoticed: recent years have seen a plethora of attempts to alleviate the burden on programmers. Several algorithmic skeleton frameworks for C++ have been extended---or purpose built---to support the orchestration of GPU computations, expressed in terms of programming patterns that leave the parallelism implicit~\cite{EnmyrenK12a,ErnstingK12a,GoliG13a,MarquesPAM13a,SteuwerG13a}. Higher-level languages on the other hand have seen new libraries, extensions, and compilers that allow for GPU programming at more comprehensible levels of abstraction, with various degrees of automatic device and memory management~\cite{DubachCRBF12a,HormatiSWMM11a,Makarov-Hauswirth15a,NystromWD11a,Pratt-SzeligaFW12a}.

These advances have made strides in the right direction, but the burden on the programmer can be lifted even further. Some approaches (e.g.~\cite{NystromWD11a}) still necessitate an understanding of relatively low-level GPU concepts such as barrier-based synchronization between threads; a mechanism that can easily lead to perplexing concurrency faults such as data races or barrier divergence. Such concepts can stifle the productivity of programmers and remain an obstacle to broadening the adoption of GPGPU. Other approaches (e.g.~\cite{DubachCRBF12a}) successfully abstract away from them, but require programmers to migrate to dedicated languages. Furthermore, to our knowledge, no existing approach has explored the possibility of integrating mechanisms or methodologies for specifying and monitoring the correctness of high-level GPU code, missing an opportunity to support the development of reliable programs. Our work has been motivated by the challenge of addressing these issues without depriving programmers of the potential performance boosts for data-parallel problems.

This paper proposes \thename{}, a library for GPU programming in the object-oriented language Eiffel, that aims to hit a sweet spot between programmer productivity on the one hand (as seen with high-level programming languages), and performance of data-parallel programs on the other (as seen with CUDA and OpenCL). First, the library binds Eiffel to the CUDA model, allowing already for GPU programs to be written, compiled, and executed by developers. Second, and crucially, \thename{} provides a high-level API for orchestrating data-parallel programs on the GPU 
that hides the low-level hardware and synchronization requirements of CUDA. Our approach aims to allow programmers to focus entirely on functionality, by offering them collections equipped with primitive data-parallel operations (e.g.~sum, max, min) that can be combined to generate complex computations, without performance becoming incommensurate with that of manually coded CUDA solutions. This is achieved by deferring the generation of CUDA kernels such that the execution of pending operations can be optimized by combining them. 

Furthermore, to support the development of safe and functionally correct GPU code, we integrate the design-by-contract~\cite{Meyer97a} methodology that is native to Eiffel; that is to say, \thename{} supports the annotation of high-level GPU programs with executable preconditions, postconditions, and invariants, together specifying the properties that should hold before and after the execution of methods. In languages supporting design-by-contract, these annotations can be checked dynamically at runtime, but the significant overhead incurred means that they are often disabled outside of debugging. With \thename{}, contracts can be constructed from the data-parallel primitives, allowing for them to be monitored at runtime with little overhead by executing them on the GPU.

The contribution of this work is hence a library for GPU programming that:

\begin{itemize}
	\item embraces the \emph{object-oriented paradigm}, shielding programmers from the low-level requirements of the CUDA model without depriving them of the performance benefits;
	
	\item is \emph{modular} and \emph{efficient}, supporting the generation of complex computations through the composition of primitive operations with a dedicated kernel optimization strategy;
	
	\item supports the writing of \emph{safe} and \emph{functionally correct code} via contracts, monitored at runtime with little overhead.
\end{itemize}

The rest of the paper is organized as follows. Section \ref{sect:overview} provides an overview of the library, its capabilities, and how it is implemented. Section \ref{sec:api} explores the binding and library APIs in more detail. Section \ref{sec:DbC_integration} describes the design-by-contract integration. Section \ref{sec:optimization} presents the kernel generation and optimization strategies. Section \ref{sec:evaluation} evaluates performance, code size, and contract checking across a selection of benchmark programs. Section \ref{sec:related} describes and contrasts some related work. In Section \ref{sec:conclusion},  we conclude.

\section{The \thename Library}
\label{sect:overview}

In this section we provide an overview of the \thename{} library. We describe the style of programming it supports for constructing GPU programs, provide a simple example, and explain how the integration with CUDA is achieved. 

\subsection{Programming Style}

CUDA kernels---the functions that run on the GPU---are executed by an array of threads, with each thread executing the same code on different data. Many computational tasks fit to this execution model very naturally (e.g.~matrix multiplication, vector addition). Many tasks, however, do not, and can only be realized with non-trivial reductions. This difficulty is compounded when one starts to write complex, multistage algorithms: combining subtasks into a larger kernel is a challenging task, and there is little support for modularity.

In contrast, \thename{} emphasizes the development of GPU programs in terms of simple, compositional ``building blocks''. For a selection of common data structures (including collections, vectors, and matrices), the library provides sets of built-in primitive operations. While individually these operations are simple and intuitive to grasp (e.g.~\code{sum}, \code{max}, \code{project}), they can also be combined and chained together to generate complex GPU computations, without the developer ever needing to think about the manipulation of kernels. The aim is to allow for developers to focus entirely on functionality, with the library itself responsible for generating kernels and applying optimizations (e.g.~combining them). This focus on functionality extends to correctness, with \thename{} supporting the annotation of programs with contracts that can be monitored efficiently at runtime.

Before we expand on these different aspects of the library, consider the simple example in Listing \ref{matrix_vector_mult}, which illustrates how a \thename program can be constructed in practice. 

\begin{Listing}
\begin{lstlisting}
matrix_transpose_vector_mult(matrix: G_MATRIX[DOUBLE]; vector: G_VECTOR[DOUBLE]): G_MATRIX[DOUBLE]
  require
    matrix.rows = vector.count
  do
    Result := matrix.transpose.right_multiply (vector)
  ensure
    Result.rows = matrix.columns
    Result.columns = 1
  end	
\end{lstlisting}
  \caption{Transposed matrix-vector multiplication example}
  \label{matrix_vector_mult}
\end{Listing}

The method takes as input a matrix and a vector, then returns the result of transposing the matrix and multiplying the vector. The computation is expressed in one line through the chaining of two compact, primitive operations from the API for matrices---\code{transpose} and \code{right_multiply}---from which the CUDA code is automatically generated and optimized. Furthermore, because the latter of the operations is only defined for inputs of certain sizes ($N \times M$ matrix; $M$ dimension vector), the method is annotated with a precondition in the \code{require} clause, expressing that the size of the input vector should be equal to the number of rows in the matrix (rows, not columns, since it will be transposed). Similarly, the postcondition in the \mbox{\code{ensure}} clause expresses the expected dimensions of the resulting matrix. Both of these properties can be monitored at runtime, with the precondition checked upon entering the method, and the postcondition checked upon exiting.

\subsection{CUDA Integration}
\label{sec:cuda_integration}

\thename provides two conceptual levels of integration with CUDA: a binding and a library level. The binding level provides a minimalistic API to run raw CUDA code within an Eiffel program, similar to bindings like PyCUDA~\cite{KlocknerPLCIF12a} and JCUDA~\cite{YanGS09a}, and is intended for experienced users who need more fine-grained control over the GPU. The library level is built on top of the binding, and provides the data structures, primitive operations, contracts, and kernel-generation facilities that form the focus of this paper.

\definecolor{lightblue}{RGB}{220,230,240}
\begin{figure}[htb]
  \centering
  \includegraphics[scale=0.97]{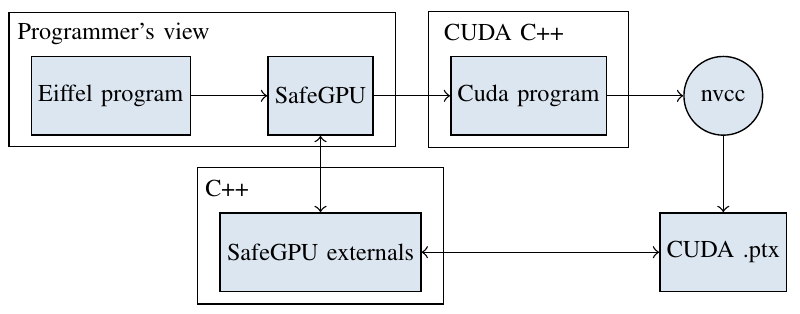}
  \caption{\thename runtime}
  \label{framework_runtime}
\end{figure}

The runtime integration of CUDA is shown in Figure~\ref{framework_runtime}. The library actively makes use of Eiffel's built-in mechanisms for interfacing with C++, allowing it to call the CUDA-specific functions it needs for initialization, data transfers, and device synchronization. These steps are handled automatically by \thename{} for both the binding and library levels, minimizing the amount of boilerplate code. Given a source kernel, whether handwritten at the binding level or generated from the library one, the \mbox{\code{nvcc}} compiler generates a \code{.ptx} file containing a CUDA module that the library can use to launch the kernel.

\section{Design of the API}
\label{sec:api}

In the following, we describe in more detail the two levels of \thename's API. First, we consider the binding, which allows expert users to run CUDA code from within Eiffel. Then we turn to the high-level library, and in particular, its three basic classes for collections, vectors, and matrices.

\subsection{CUDA Binding}
\label{sec:binding_api}

The binding API provides handles to access the GPU and raw memory. Programming with this API requires effort comparable to plain CUDA solutions and is therefore not a user-level API; its main purpose is to provide functionality for the library API built on top of it.

\tabref{binding_api_table} provides details about the API's classes. The two main classes are \code{CUDA_KERNEL} and \code{CUDA_DATA_HANDLE}. The former encapsulates a CUDA kernel; the latter represents a
contiguous sequence of uniform objects, e.g.~a single-dimensional array. 

\begin{table}[h]
\caption{Overview of the binding API}
\label{tab:binding_api_table}
\vspace{-1.5ex}
\centering
\renewcommand{\arraystretch}{1.2}
{\small
\begin{tabular}{lp{5.2cm}}
class                    & description   \\ \hline
\scode{CUDA_DATA_HANDLE} & Represents a handle to a device memory location. Supports scalar, vector, and multi-dimensional data. Can be created from (and converted to) standard \scode{ARRAY}s. \\
\scode{CUDA_INTEROP}     & Encapsulates low-level device operations, such as initialization, memory allocation, and data transfer. \\
\scode{CUDA_KERNEL}      & Represents a CUDA kernel, ready for execution. Can contain an arbitrary number of \scode{CUDA_DATA_HANDLE} kernel inputs, one of which is used as output. Can be launched with configurable shared memory. \\
\scode{LAUNCH_PARAMS}    & Encapsulates the grid setup and shared memory size required to launch a \scode{CUDA_KERNEL}. \\
\scode{KERNEL_LOADER}    & Is responsible for loading CUDA kernels into the calling process. If necessary, performs a kernel compilation. Can load kernels from a file or from a string. \\
\end{tabular}
}
\end{table}

\subsection{Collections}
\label{sec:collections}

Collections are the most abstract container type provided by \thename; the majority of bulk operations---operating on an entire collection---are defined here. Collections are array-based, i.e.~they have bounded capacity and count, and their items can be accessed by index. Collections do not automatically resize, but new ones with different sizes can be created using the methods of the class.

The key methods of the collection API are given in \tabref{collection_api_table} and described in the following (in Eiffel, \code{like Current} denotes the type of the current object). A \thename collection can be created using the method \mbox{\code{from_array},} which creates its content from that of an Eiffel array: as an array's content is contiguous, a single call to CUDA's analogue of \code{memcpy} suffices. Individual elements of the collection can then be accessed through the method \code{item}, and the total number of elements is returned by \mbox{\code{count}.} The method \code{concatenate} is used to join the elements of two containers and the method \code{subset} resizes a given collection to a subset.

 \begin{table}[htb]
\caption{Collection API}
 \label{tab:collection_api_table}
\vspace{-2ex}
 {\small
 \begin{description}
 \lstitem{from_array (array: ARRAY[T])}
   Creates an instance of a collection, containing items from the standard Eiffel array provided as input.
 \lstitem{item (i: INT): T} 
   Access to a single element. 
 \lstitem{count: INT}
   Queries the number of elements in the collection.
 \lstitem{concatenate (other: like Current): like Current}
   Creates a new container consisting of the elements in the current object followed by those in \code{other}.
 \lstitem{subset(start, finish: INT): like Current}  
   Creates a subset of the collection that shares the same memory as the original.
 \lstitem{for_each (action: PROCEDURE[T]): like Current}  
   Applies the provided procedure to every element of the collection.
 \lstitem{project (transform: FUNCTION[T, U]): COLLECTION[U]}  
   Performs a projection operation on the collection: each element is transformed according to the specified function.
 \lstitem{filter (condition: PREDICATE[T]): like Current}
   Creates a new collection containing only items for which the specified predicate holds.
 \lstitem{for_all (condition: PREDICATE[T]): BOOLEAN}  
   Checks whether the specified predicate holds for all items in the collection.
 \lstitem{exists (condition: PREDICATE[T]): BOOLEAN}  
   Checks whether the specified predicate holds for at least one item in the collection.
 \lstitem{new_cursor: ITERATION_CURSOR [T]}   
   Implementation of \scode{ITERABLE[T]}; called upon an iteration over the collection.
 \lstitem{update}
   Forces execution of all pending operations associated with the current collection. The execution is optimized whenever possible.
 \end{description}
 }
  \end{table}

The core part of the API design consists of methods for transforming, filtering, and querying collections. All these methods make use of Eiffel's functional capabilities in the form of \emph{agents}, which represent operations (similar to delegates in C\# or anonymous classes in Java) that are applied in different ways to all the elements of a collection. Agents can be one of three types: \emph{procedures}, which express transformations to be applied to elements (but do not return results); \emph{functions}, which return results for elements (but unlike procedures, are side-effect free); or \emph{predicates}, which are Boolean expressions.

To construct a new collection from an existing one, the API provides the transformation methods \code{for_each} and \code{project}. The former applies a procedure agent to each element of the collection, whereas the latter applies a function agent. For example, the call

\vspace{1ex}
\lstinline[basicstyle=\small,mathescape]{c.for_each(agent (a: INT) do a := a * 2 end)} 
\vspace{1ex}

\noindent represents an application of \code{for_each} to an integer collection \code{c}, customized with a procedure that doubles every element. In contrast, the call

\vspace{1ex}
\lstinline[basicstyle=\small,mathescape]{c.project(agent (a: INT): DOUBLE do Result := sqrt(a) end)}
\vspace{1ex}

\noindent creates from an integer collection \code{c} a collection of doubles, with each element the square root of the corresponding one in \code{c}.

To filter or query a collection, the API provides the methods \mbox{\code{filter},} \mbox{\code{for_all},} and \mbox{\code{exists}}, which evaluate predicate agents with respect to every element. An example of filtering is

\vspace{1ex}
\lstinline[basicstyle=\small,mathescape]{c.filter(agent (a: INT) do Result := a < 5 end)} 
\vspace{1ex}

\noindent which creates a new collection from an integer collection \mbox{\code{c},} containing only the elements that are less than five. The method \code{for_all} on the other hand does not create a new collection, but rather checks whether the predicate holds for every element or not; the call

\vspace{1ex}
\lstinline[basicstyle=\small,mathescape]{c.for_all(agent (i: T) do Result := pred(i) end)}
\vspace{1ex}

\noindent returns \code{True}, for example, if some (unspecified) predicate \code{pred} holds for every element of the collection \code{c} (and \code{False} otherwise). The method \code{exists} is similar, returning \code{True} if the predicate holds for at least one element in the collection (and \code{False} otherwise).

The queries \code{for_all} and \code{exists} are particularly useful in contracts, and can be parallelized effectively for execution on the GPU. We discuss this further in Section \ref{sec:DbC_integration}.

Collections are embedded into Eiffel's container hierarchy by implementing the \code{ITERABLE} interface, which allows the enumeration of their elements in foreach-style loops (\code{across} in Eiffel terminology). Enumerating is efficient: upon a call to \code{new_cursor}, the collection's content is copied back to main memory in a single action. 

Finally, the special method \code{update} forces execution of any pending kernel operations (described further in Section \ref{sec:optimization}).

\subsection{Vectors}

Vectors are a natural specialization of collections. Besides the capabilities of collections, they provide a range of numerical operations.

The API for vectors 
allows for computing the average value \code{avg} and \code{sum} of the elements of arbitrary vectors, as well as computing the minimal \code{min} and maximal \code{max} elements. Furthermore, \code{is_sorted}will check whether the elements are sorted. These functions are all implemented by multiple reductions on the device side; the computation is optimal in the sense that computation via reduction does not do more work than its sequential counterpart.

All numerical operations such as \code{plus} and \code{minus} (alongside in-place variants), as well as \code{multiplied_by} and \code{divided_by} (alongside component-wise variants) are defined as vector operations on the GPU, e.g.~a call to \code{plus} performs vector addition in a single action on the device side. Note that aliases can be used for value-returning operations, e.g.~\code{v * n} instead of \mbox{\code{v.multiplied_by(n)}}.

An important requirement in using and composing vector operations is keeping the dimensions of the data synchronized. Furthermore, certain arithmetic operations are undefined on certain elements; \mbox{\code{divided_by},} for example, requires that elements are non-zero. Such issues are managed through contracts built-in to the API that can be monitored at runtime, shielding developers from inconsistencies. We discuss this further in Section \ref{sec:DbC_integration}.

\subsection{Matrices}

The matrix API is strongly tied to the vector API: the class uses vectors to represent rows and columns. On the device side, a matrix is stored as a single-dimensional array with row-wise alignment. Thus, a vector handle for a row can be created by adjusting the corresponding indices. The column access pattern is more complicated, and is implemented by performing a copy of corresponding elements into new storage.

In the API, the queries \code{rows} and \code{columns} return the dimensions of the matrix, whereas \code{item}, \code{row}, and \code{column} return the parts of the matrix specified. Single-column or single-row matrices can be converted to vectors by making appropriate calls to the \code{row} or \code{column} methods.

Similar to vectors, the API provides both conventional and in-place methods for addition and subtraction. Beyond these primitive arithmetic operations, the API provides built-in support for matrix-matrix multiplication (method \mbox{\code{multiply})} since it is a frequently occurring operation in GPGPU. The implementation optimizes performance through use of the shared device memory.

The API also supports scalar multiplication \mbox{(\code{multiplied_by})}, left and right matrix-vector multiplication (\code{left_multiply} and \code{right_multiply}), component-wise matrix multiplication and division \mbox{(\code{compwise_multiply}} and \code{compwise_divide}), matrix transposition (\code{transpose}), and \code{submatrix} creation.

Similar to the other API classes, matrix methods are equipped with contracts in order to shield the programmer from common errors, e.g.~mismatching dimensions in matrix multiplication.

\section{Design-by-Contract Integration}
\label{sec:DbC_integration}

To support the development of safe and functionally correct code, \thename integrates the design-by-contract~\cite{Meyer97a} methodology native to the Eiffel language, i.e.~the annotation of methods with executable pre- and postconditions, expressing precisely the properties that should hold upon entry and exit. These can be monitored at runtime to help ensure the correctness of programs. In the context of GPU programs, in which very large amounts of data might be processed, ``classical'' (i.e.~sequential) contracts take so long to evaluate that they need to be disabled outside of debugging. With \thename{}, however, contracts can be expressed using the primitive operations of the library itself, and thus can be executed on the GPU---where the data is sitting---without diminishing the performance of the program (see our benchmarks in \secref{contract_perfomance}).

Contracts are utilized by \thename programs in two ways. First, they are built-in to the library API; several of its methods are equipped with pre- and postconditions, providing correctness properties that can be monitored at runtime ``for free'' (i.e.~without requiring additional user annotations). Second, when composing the methods of the API to generate more complex, compound computations, users can define and thus monitor their own contracts expressing the intended effects of the overall computation.

The API's built-in contracts are easily motivated by vector and matrix mathematics, for which several operations are undefined on input with inconsistent dimensions or input containing zeroes. Consider for example \lstref{vector_contract_example}, which contains the signature and contracts of the library method for component-wise vector division. Calling \code{v1.compwise_divide(v2)} on vectors \code{v1} and \code{v2} of equal size results in a new vector, constructed from \code{v1} by dividing its elements by the corresponding elements in \code{v2}. The preconditions in the \code{require} clause assert that the vectors are of equal size (via \code{count}, from the collection API) and that all elements of the second vector are non-zero (via \code{for_all}, customized with a predicate agent). The postcondition in the \code{ensure} clause characterizes the effect of the method by asserting the expected relationship between the resulting vector and the input (retrieved using the \code{old} keyword).

\begin{Listing}
\begin{lstlisting}
compwise_divide(other: VECTOR[T]): VECTOR[T]
	require
		other.count = count
		other.for_all(
			agent (el: T) do Result := el /= {T}.zero end)
	ensure
		Current = old Current
		Result * other = Current 
	end
\end{lstlisting}
  \caption{Contracts for component-wise vector division}
  \label{vector_contract_example}
\end{Listing}

\begin{Listing*}
\begin{lstlisting}
quicksort(a: G_VECTOR[REAL_32]): G_VECTOR[REAL_32]
  require
    a.count > 0
  local
    pivot: DOUBLE
    left, mid, right: G_VECTOR[REAL_32]
  do
    if (a.count = 1) then
      Result := a
    else
      pivot := a[a.count // 2]

      left := a.filter (agent (item: REAL_32; a_pivot: REAL_32): BOOLEAN do Result := item < a_pivot end (?, pivot))
      right := a.filter (agent (item: REAL_32; a_pivot: REAL_32): BOOLEAN do Result := item > a_pivot end (?, pivot))
      mid := a.filter (agent (item: REAL_32; a_pivot: REAL_32): BOOLEAN do Result := item = a_pivot end (?, pivot))

      Result := quicksort(left).concatenate(mid).concatenate(quicksort(right))
    end
  ensure
    Result.is_sorted
    Result.count = a.count
  end
\end{lstlisting}
  \caption{\thename implementation of quicksort}
  \label{sort_contracts}
\end{Listing*}

\thename{} provides a straightforward way to monitor user-defined contracts on the GPU: simply express them in the \code{require} and \code{ensure} clauses of methods, using the primitive operations of the library (this is analogous to classical design-by-contract, in which methods are used in both specifications and implementations). Consider for example the implementation and contracts of \code{quicksort} in Listing~\ref{sort_contracts}. The implementation utilizes two methods provided by the collection API: \code{concatenate}, to efficiently concatenate two vectors; and \code{filter}, to find items less than, greater than, or equal to the pivot. The three calls to \code{filter} are customized with predicate agents expressing these relations. Since inline agents cannot access local variables in Eiffel, the \code{pivot} is passed as an argument. This is denoted by \code{(?, pivot)} at the end of each agent expression; here, the \code{?} corresponds to \code{item}, and expresses that it should be instantiated with successive elements of the collection; \code{pivot} corresponds to \code{a_pivot}, and expresses that the latter should always take the value of the former. At runtime, the built-in contracts of these two library methods can be monitored, but they only express correctness conditions localized to their use, and nothing about their compound effects. The overall postcondition of the computation can be expressed as a user-defined postcondition of \code{quicksort}, here asserting---using the \code{is_sorted} and \code{count} methods of the vector API---that the resulting vector is sorted and of the same size. This can be monitored at runtime to increase confidence that the user-defined computation is correct.

Built-in and user-defined contracts for GPU collections are typically classified as one of two types. \emph{Scalar contracts} are those using methods with evaluation times independent of the collection size. A common example is \code{count}, which records the number of elements a collection contains. \emph{Range contracts} are those using methods that operate on the elements of a collection with evaluation times that grow with the collection size. These include library methods such as \code{sum}, \code{min}, \code{max}, and \code{is_sorted}; the CUDA programs generated for such operations usually perform multiple reductions on the GPU. Other common range contracts are those built from \code{for_all} and \code{exists}, equipped with predicate agents, expressing properties that should hold for every (resp.~at least one) element of a collection. These are easily parallelized for execution on the GPU, and unlike their sequential counterparts, can be monitored at runtime for very large volumes of data without diminishing the overall performance of the program (see \secref{contract_perfomance}).

\section{Kernel Generation and Optimization}
\label{sec:optimization}

In this section we describe how \thename{} translates individual methods of the API to CUDA kernels, how data is managed, and how the library optimizes kernels for compound computations.

Generating CUDA kernels for calls of individual library methods is straightforward. Each method is associated with a kernel template, which the library instantiates with respect to the particular collection and parameters of the method call. The \thename{} runtime (as described in Section \ref{sec:cuda_integration}) then handles its execution on the GPU via Eiffel's mechanisms for interfacing with C++.

Transferring data to and from the GPU is expensive, so the library attempts to minimize the number of occurrences. The only time that data is transferred to the GPU is upon calling the method \mbox{\code{from_array},} which creates a GPU collection from a standard Eiffel array. Once the data is on the GPU, it remains there for arbitrarily many kernels to manipulate and query (including those corresponding to contracts). Operations that create new collections from existing ones (e.g.~\mbox{\code{filter},} \mbox{\code{project})} do so without transferring data away from the GPU; this occurs only for methods that specifically query them.

While the primitive operations alone already support many useful computations (e.g.~matrix multiplication, vector addition), the heart of \thename{} is in its support for combining and chaining such operations to implement multistage algorithms on the GPU. The main challenge for a library aiming to provide this support is to do so without performance becoming incommensurate with that of manually written CUDA kernels. A naive solution, for example, might have been to generate one kernel per method call and launch them one after the other. With \thename{} however, we adopt a deferred execution model, analyze pending kernels, and attempt to generate more efficient CUDA code by combining them.

By default, a method call is not executed, but rather added to a list of pending actions for the corresponding collection. There are three ways to trigger its execution: (1) perform a function call that returns a scalar value, e.g.~\code{sum}; (2) perform a call to \code{to_array} which creates a standard Eiffel array from the GPU collection; or (3) perform a call of the special method \code{update}, which forces the execution of any pending kernels.

Consider for example the problem of computing the dot product (or inner product) of two vectors, which can be solved by combining vector multiplication and vector summation as in Listing~\ref{dot_product}. Here, the result is obtained by chaining the \mbox{\code{a.compwise_multiply (b)}} method---which produces an anonymous intermediate result---with \mbox{\code {vector.sum}.} In this example, the computation is deferred until the call of \code{sum}, which returns the sum of the elements in the vector.

\begin{Listing}
\begin{lstlisting}
dot_product(a, b: G_VECTOR[DOUBLE]): DOUBLE
  require
    a.count = b.count
  do
    Result := a.compwise_multiply (b).sum 
    -- component-wise vector multiplication, followed by sum of its elements
  end
\end{lstlisting}
  \caption{Combining primitives to compute the dot product}
  \label{dot_product}
\end{Listing}

\begin{figure}[h]
  \centering
\begin{subfigure}[t]{0.55\columnwidth}
  \centering
{\scriptsize
\begin{tikzpicture}[nodes={draw},node distance=3ex,minimum size=.75cm,inner sep=0pt,yscale=0.5]
\node [rectangle] (ABk) {cmult};
\node [circle] (A) [above left=of ABk] {$X$};
\node [circle] (B) [below left=of ABk] {$Y$};
\node [circle] (C) [right=of ABk] {$XY$};
\node [rectangle] (Csum) [right=of C] {sum};
\node [circle] (Dot) [right=of Csum] {$X\! \cdot\! Y$};

\path[->] 
(A) edge (ABk) 
(B) edge (ABk)
(ABk) edge (C)
(C) edge (Csum)
(Csum) edge (Dot);
\end{tikzpicture}
}
\vspace{-2.5ex}
\caption{Before optimization}
\label{fig:dotbefore}
\end{subfigure}
\hfill
\begin{subfigure}[t]{0.35\columnwidth}
  \centering
{\scriptsize
\begin{tikzpicture}[nodes={draw},node distance=3ex,minimum size=.75cm,inner sep=0pt]
\node [rectangle] (ABk) {\begin{tabular}{@{}c@{}}cmult \\ \hline sum\end{tabular}};
\node [circle] (A) [above left=of ABk] {$X$};
\node [circle] (B) [below left=of ABk] {$Y$};
\node [circle] (Dot) [right=of ABk] {$X\! \cdot\! Y$};

\path[->] 
(A) edge (ABk) 
(B) edge (ABk)
(ABk) edge (Dot);
\end{tikzpicture}
}
\caption{After optimization}
\label{fig:dotafter}
\end{subfigure}
  \caption{Execution plans for the dot product method}
  \label{optimizing_dot_product}
\end{figure}
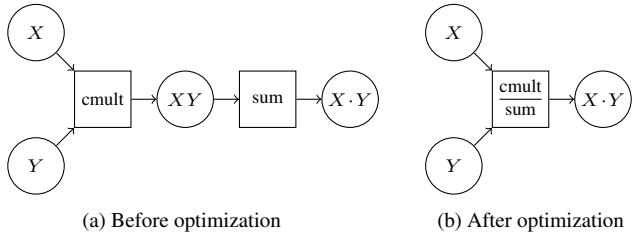

The benefit of deferring execution until necessary is that the kernel code can be optimized. Instead of generating kernels for every method call, \thename{} uses some simple strategies to merge deferred calls and thus handle the combined computation in fewer kernels. Before generating kernels, the optimizer constructs an execution plan from the pending operations. The plan takes the form of a DAG, representing data and kernels as two different types of nodes, and representing dependencies as edges between them. The optimizer then traverses the DAG, merging kernel vertices and collapsing intermediate dependencies where possible. Upon termination, the kernel generation takes place on the basis of the optimized DAG. We illustrate a typical optimization in Figure~\ref{optimizing_dot_product}, which shows the execution plans for the dot product method of Listing~\ref{dot_product}.

The plan in Figure~\ref{fig:dotbefore} is the original one extracted from the pending operations; this would generate two separate kernels for multiplication and summation (cmult and sum) and launch them sequentially. The plan in Figure~\ref{fig:dotafter}, however, is the result of an optimization; here, the deferred cmult kernel is combined with sum. The combined kernel generated by this optimized execution plan would perform component-wise vector multiplication first, followed by summation, with the two stages separated using barrier synchronization. This simple optimization pattern extends to several other similar cases in \thename{}.

\begin{figure}[htb]
  \centering
\begin{subfigure}[b]{0.55\columnwidth}
  \centering
{\scriptsize
\begin{tikzpicture}[nodes={draw},node distance=3ex,minimum size=.75cm,inner sep=0pt,yscale=0.5]
\node [rectangle] (aXk) {smult};
\node [circle] (a) [above left=of ABk] {$a$};
\node [circle] (X) [below left=of ABk] {$X$};
\node [circle] (aX) [right=of ABk] {$aX$};
\node [rectangle] (XY) [above right=of aX] {plus};
\node [circle] (Y) [above left=of XY] {$Y$};
\node [circle] (AxY) [right=of XY] {$aX\!+\!Y$};

\path[->] 
(a) edge (aXk) 
(X) edge (aXk)
(aXk) edge (aX)
(aX) edge (XY)  
(Y) edge (XY)  
(XY) edge (AxY);
\end{tikzpicture}
}
\vspace{-2.5ex}
\caption{Before optimization}
\end{subfigure}
\hfill
\begin{subfigure}[b]{0.35\columnwidth}
  \centering
{\scriptsize
\begin{tikzpicture}[nodes={draw},node distance=3ex,minimum size=.75cm,inner sep=0pt]
\node [rectangle] (ABk) {\begin{tabular}{@{}c@{}}smult \\ \hline plus\end{tabular}};
\node [circle] (A) [above left=of ABk] {$Y$};
\node [circle] (B) [below left=of ABk] {$X$};
\node [circle] (C) [left=of ABk] {$a$};
\node [circle] (Dot) [right=of ABk] {$aX\!+\!Y$};

\path[->] 
(A) edge (ABk) 
(B) edge (ABk)
(C) edge (ABk)
(ABk) edge (Dot);
\end{tikzpicture}
}
\caption{After optimization}
\end{subfigure}
  \caption{Execution plans for vector mathematics}
  \label{linear_optimization}
\end{figure}
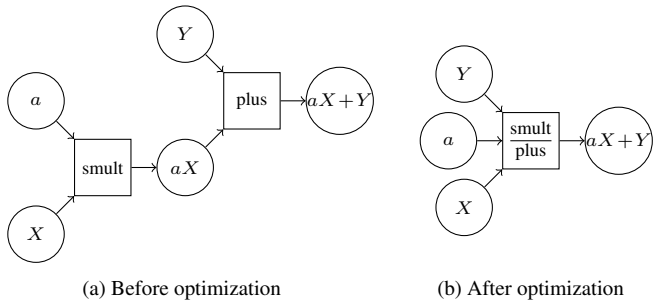

The optimizer is particularly well-tuned for computations involving vector mathematics. In some cases, barriers are not needed at all; the optimizer simply modifies the main expression in the kernel body, leading to more efficient code. For example, to compute $aX + Y$ where $a$ is a scalar value and $X$, $Y$ are vectors, the optimizer just slightly adjusts the vector addition kernel, replacing \code{X[i] + Y[i]} with \code{a*X[i] + Y[i]}. Such optimizations also change the number of kernel arguments, as shown in Figure~\ref{linear_optimization}.

Listing~\ref{gaussian_elimination} shows the usefulness of the optimizer as part of an extended example (Gaussian elimination to find the determinant of a matrix). Here, in the inner loop of the Gaussian elimination, the operations

\vspace{1ex}
\lstinline[basicstyle=\small,mathescape]{matrix.row(i).divided_by (pivot)}

\lstinline[basicstyle=\small,mathescape]{matrix.row(i).in_place_minus(matrix.row (step))}
\vspace{1ex}

\noindent are combined by the optimizer to generate \code{(A[i] / pivot) - A[step]} in the kernel.

\begin{Listing}
\begin{lstlisting}
gauss_determinant (matrix: G_MATRIX[DOUBLE]): DOUBLE
  require
    matrix.rows = matrix.columns
  local
    step, i: INTEGER
    pivot: DOUBLE
  do
    Result := 1
    from
      step := 0
    until
      step = matrix.rows
    loop
      pivot := matrix(step, step)
      Result := Result * pivot

      if not double_approx_equals (pivot, 0.0) then
        matrix.row (step).divided_by (pivot)
      else
        step := matrix.rows
      end
      from
        i := step + 1
      until
        i = matrix.rows
      loop
        pivot := matrix(i, step)
        if not double_approx_equals (pivot, 0.0) then
          matrix.row(i).divided_by (pivot)
          matrix.row(i).in_place_minus(matrix.row (step))
        end
        i := i + 1
      end

      step := step + 1
    end
  end
\end{lstlisting}
  \caption{\thename implementation of Gaussian elimination}
  \label{gaussian_elimination}
\end{Listing}

\section{Evaluation}
\label{sec:evaluation}

To evaluate \thename{}, we designed a set of benchmark GPU programs encompassing problems that fit naturally to the execution model (e.g.~vector addition, matrix multiplication), as well as more general-purpose ones constructed by chaining the primitive operations of the library (e.g.~Gaussian elimination, quicksort). Across these benchmarks we made three different comparisons:
\begin{enumerate}
\item the performance of the high-level API against corresponding CUDA and Eiffel implementations;
\item the conciseness of functionally equivalent programs in \thename{} and sequential Eiffel;
\item the performance overhead of runtime contract checking, compared with checking traditional sequential contracts.
\end{enumerate}

The six benchmark programs we considered were vector addition, dot product, matrix multiplication, Gaussian elimination, quicksort, and matrix transposition. Each benchmark was implemented using \thename{} (with contracts for the GPU, wherever possible) and traditional Eiffel (with sequential contracts, wherever possible). We did not implement but rather relied on a selection of sources for the plain CUDA implementations: vector addition and matrix multiplication were taken from the NVIDIA SDK, with dot product and quicksort adapted from code in the same source; Gaussian elimination came from a parallel computing research project~\cite{CUDA-Gaussian}; and finally, matrix transposition came from a post~\cite{CUDA-Transpose} on NVIDIA's Parallel Forall blog.

All experiments were performed on the following hardware: Intel Core i7 8 cores, 2.7 GHz; NVIDIA QUADRO K2000M (2 GB memory, compute capability 3.0). In our measurements we are reporting wall time. Furthermore, we measure only the relevant part of the computation, omitting the time it takes to generate the input.

The \thename{} implementations of quicksort and Gaussian elimination are provided in this paper (Listings~\ref{sort_contracts} and~\ref{gaussian_elimination} respectively). Our source code and the other benchmarks are available online\footnote{See: \url{https://bitbucket.org/alexey_se/eiffel2cuda}}.\\

\subsection{Performance}
\label{sec:performance:performance}

The main goal of our first experiment was to compare the performance of \thename{} programs against plain CUDA implementations with the same functionality, in order to assess what the higher level of abstraction translates to in terms of performance overhead. We also recorded the performance of functionally equivalent programs in sequential Eiffel, to determine the sizes of inputs for which the GPU approaches start to outperform it. We remark that for this experiment, contract checking was completely disabled.

The results of our comparison are shown in Figure \ref{fig:perf_table1}. The problem size ($x$-axis) is defined for both vectors and matrices as the total number of elements they contain (our benchmarks use only square matrices, hence the number of rows or columns is always the square root). The times ($y$-axis) are shown in seconds, and are the medians of ten runs.

While sequential Eiffel is faster than \thename{} and plain CUDA on relatively small inputs (as expected, because of the overhead from launching the GPU), it is outperformed by both when the size of the data becomes large. This happens particularly quickly for the non-linear algorithm (e) in comparison to the others.

The results provide support for our argument that using our library does not lead to performance incommensurate with that of handwritten CUDA code: its performance is very close to that of plain CUDA across most of the benchmarks. The Gaussian elimination benchmark (d) is an exception for larger inputs. This is due to the need for the \thename{} implementation to use loops, which have the effect of additional kernel launches in comparison to the handwritten CUDA code. In other benchmarks, \thename{} sometimes slightly outperforms plain CUDA, which could be due to differences between the memory managers of Eiffel and C++.

\begin{figure*}[t]
   	\centering
\pgfplotsset{every axis/.append style={font=\scriptsize}}
	\begin{subfigure}[b]{0.3\textwidth}
        \centering
        \begin{tikzpicture}
			\begin{loglogaxis}[xtick={1e3, 1e4, 1e5, 1e6, 1e7,1e8},ylabel=Time (seconds),height=5cm,width=6.2cm, minor tick length=0,y label style={yshift=-2ex}]
				\addplot[mark=square*,color=blue] 
					coordinates {
						(1e3,0.102)
						(1e4,0.1)
						(1e5,0.102)
						(1e6,0.1035)
						(1e7,0.1425)
						(1e8,0.5475)
					};
								
				\addplot[mark=diamond*,color=red] 
					coordinates {
						(1e3,0.0001)
						(1e4,0.0007)
						(1e5,0.003)
						(1e6,0.027)
						(1e7,0.2665)
						(1e8,2.6685)
					};

				\addplot[mark=triangle*,color=green] 
					coordinates {
						(1e3,0.1025)
						(1e4,0.1005)
						(1e5,0.101)
						(1e6,0.103)
						(1e7,0.125)
						(1e8,0.4035)
					};
			\end{loglogaxis}
		\end{tikzpicture}
		\caption{Vector Addition}
		\label{fig:vector-addition}
		\end{subfigure}
	\hfill
	\begin{subfigure}[b]{0.3\textwidth}
        \centering
        \begin{tikzpicture}
			\begin{loglogaxis}[xtick={1e3, 1e4, 1e5, 1e6, 1e7,1e8},height=5cm,width=6.2cm, minor tick length=0]
				\addplot[mark=square*,color=blue] 
					coordinates {
						(1e3,0.1005)
						(1e4,0.098)
						(1e5,0.0995)
						(1e6,0.1015)
						(1e7,0.1445)
						(1e8,0.546)
					};

				\addplot[mark=diamond*,color=red] 
					coordinates {
						(1e3,0.0001)
						(1e4,0.0008)
						(1e5,0.002)
						(1e6,0.02)
						(1e7,0.1975)
						(1e8,1.9845)
					};

				\addplot[mark=triangle*,color=green] 
					coordinates {
						(1e3,0.0965)
						(1e4,0.0985)
						(1e5,0.106)
						(1e6,0.0995)
						(1e7,0.1385)
						(1e8,0.47)
					};
					
			\end{loglogaxis}
		\end{tikzpicture}
		\caption{Dot Product}
		\label{fig:dot-product}
		\end{subfigure}
\hfill
\begin{subfigure}[b]{0.3\textwidth}
        \centering
        \begin{tikzpicture}
			\begin{loglogaxis}[xtick={1e3, 1e4, 1e5, 250000, 1e6},height=5cm,width=6.2cm, minor tick length=0]
				\addplot[mark=square*,color=blue] 
					coordinates {
						(1e3,0.089)
						(1e4,0.0915)
						(1.1*1e5,0.0915)
						(2.5*1e5,0.099)
						(1e6,0.1985)
					};
					
				\addplot[mark=diamond*,color=red] 
					coordinates {
						(1e3,0.001)
						(1e4,0.022)
						(1.1*1e5,0.7995000000000001)
						(2.5*1e5,2.7035)
						(1e6,21.8175)
					};
				\addplot[mark=triangle*,color=green] 
					coordinates {
						(1e3,0.08)
						(1e4,0.0815)
						(1.1*1e5,0.095)
						(2.5*1e5,0.118)
						(1e6,0.3525)
					};
					
			\end{loglogaxis}
		\end{tikzpicture}
		\caption{Matrix-Matrix Multiplication}
		\label{fig:mat-mul}
		\end{subfigure}
~\\[2ex]
\begin{subfigure}[b]{0.3\textwidth}
        \centering
        \begin{tikzpicture}
			\begin{loglogaxis}[xtick={1e3, 1e4, 1e5, 250000, 1e6},ylabel=Time (seconds),height=5cm,width=6.2cm, minor tick length=0, y label style={yshift=-2ex}]
				\addplot[mark=square*,color=blue]
					coordinates { 
						(1e3,	0.093)
						(1e4,	0.102)
						(1.1*1e5,	0.143)
						(2.5*1e5,	0.188)
						(1e6,	0.2521)
					};
				\addplot[mark=diamond*,color=red] 
					coordinates {
						(1e3,	0.001)
						(1e4,	0.012)
						(1.1*1e5,	0.403)
						(2.5*1e5,	1.351)
						(1e6,	10.7595)
					};
				\addplot[mark=triangle*,color=green]
					coordinates { 
						(1e3,	0.089)
						(1e4,	0.112)
						(1.1*1e5,	0.139)
						(2.5*1e5,	0.217)
						(1e6,	0.752)
					};
			\end{loglogaxis}
	\end{tikzpicture}
	\caption{Gaussian Elimination}
\label{fig:gaussian-elimination}
\end{subfigure}
\hfill
\begin{subfigure}[b]{0.3\textwidth}
        \centering
        \begin{tikzpicture}
\begin{loglogaxis}[xtick={1e3, 1e4, 1e5, 1e6, 1e7,1e8},height=5cm,width=6.2cm, minor tick length=0,
legend style={
overlay,
at={(-0.55,1.65)},
anchor=center}]

\legend{CUDA, Eiffel, \thename}

\addplot[mark=square*,color=blue]
coordinates {
(1e3, 0.089)
(1e4, 0.09)
(1e5, 0.091)
(1e6, 0.250)
(1e7, 0.914)
(1e8, 1.677)
};

\addplot[mark=diamond*,color=red]
coordinates {
(1e3, 0.007)
(1e4, 0.077)
(1e5, 0.6315)
(1e6, 5.0504999999999995)
(1e7, 48.9875)
(1e8, 469.0385)

};

\addplot[mark=triangle*,color=green]
coordinates {
(1e3, 0.081)
(1e4, 0.091)
(1e5, 0.0999)
(1e6, 0.361)
(1e7, 1.451)
(1e8, 2.4)
};
\end{loglogaxis}
\end{tikzpicture}
\caption{Quicksort}
\label{fig:quicksort}
\end{subfigure}
\hfill
\begin{subfigure}[b]{0.3\textwidth}
        \centering
        \begin{tikzpicture}
			\begin{loglogaxis}[xtick={1e3, 1e4, 1e5, 1e6, 1e7, 1e8},height=5cm,width=6.2cm, minor tick length=0]
				\addplot[mark=square*,color=blue] 
					coordinates {
						(1e3,0.08)
						(1e4,0.082)
						(1.1*1e5,0.082)
						(1e6,0.086)
						(1e7, 0.093)
						(1e8, 0.133)
					};
					
				\addplot[mark=diamond*,color=red]  
					coordinates {
						(1e3,0.001)
						(1e4,0.002)
						(1.1*1e5,0.006)
						(1e6, 0.03)
						(1e7, 0.4045)																		
						(1e8, 4.691000000000001)
					};
				\addplot[mark=triangle*,color=green] 
					coordinates {
						(1e3,0.08)
						(1e4,0.082)
						(1.1*1e5,0.083)
						(1e6,0.088)
						(1e7, 0.096)
						(1e8, 0.14)
					};
					
			\end{loglogaxis}
		\end{tikzpicture}
		\caption{Matrix Transpose}
		\label{fig:mat-tr}
		\end{subfigure}
\hfill
\begin{subfigure}[b]{0.3\textwidth}
        \centering
        \begin{tikzpicture}
\end{tikzpicture}
\end{subfigure}

   \vspace{-2ex}
   \caption{\thename performance evaluation}
   \label{fig:perf_table1}
\end{figure*}

\subsection{Code Size}

The second part of our evaluation considers code size, and in particular, lines of code (LOC). We only compare the \thename{} and sequential Eiffel programs in this part, and not the plain CUDA code. This is because it is not a particularly interesting comparison to make: it is known that higher-level languages are usually more compact than those at the C/C++ level of abstraction~\cite{nanz-furia:2015:rosetta}, and CUDA programs in particular are dominated by explicit memory management that is not visible in \thename{} or Eiffel.

Our results are presented in \tabref{loc_table}. The programs written using our library are quite concise (as expected for a high-level API), but more surprisingly, are more compact than the traditional sequential Eiffel programs. This is explained by the usage of looping constructs. In sequential Eiffel, loops are frequently used to implement the benchmarks. With \thename{} however, loops are often avoided due to the presence of bulk operations in the API, i.e.~operations that apply agents to all the data present in a collection. We should note that this is not always the case, as loops were required to implement the library version of the Gaussian elimination benchmark.

\begin{table}[h]
\caption{LOC comparison}
\label{tab:loc_table}
\vspace{-1ex}
	\centering
{\normalsize
\begin{tabular}{lrrrr}
problem                       & Eiffel      & \thename & ratio \\ \hline
Vector Addition               &   15        & 6  &  2.5 \\
Dot Product                   &   17        & 7  &  2.4 \\
Matrix-Matrix Multiplication  &   31        & 9  &  3.4 \\
Gaussian Elimination          &   97        & 34 &  2.9 \\
Quicksort                     &  110        & 64 &  1.7 \\
Matrix Transpose              &  26        & 8   &  3.3 \\
\end{tabular}
}
\end{table}

\subsection{Contract Overhead}
\label{sec:contract_perfomance}

The goal of our final experiment was to measure the cost of checking \thename{} contracts on the GPU against the cost of checking traditional sequential Eiffel ones. To allow a more fine-grained comparison, we measured the contract checking overhead in three different modes: (1) preconditions only; (2) pre- and postconditions only; and finally, (3) full contract checking, i.e.~additionally checking class invariants at method entry and exit points. Our benchmarks were annotated only with pre- and postconditions; invariants, however, are present in the core Eiffel libraries that were required to implement the sequential programs (these libraries also include some additional pre- and postconditions, making a full like-for-like comparison with \thename{} challenging). Across the benchmarks and for increasingly large sizes of input, we computed ratios expressing the performance overhead resulting from enabling each of these three modes against no contract checking at all. The ratios are based on medians of ten runs (an effect of this is that some ratios can be less than 1).

Our data is presented in \tabref{contract_overhead_comparison}, where a ratio $X$ can be interpreted as meaning that the program was $X$ times slower with the given contract checking mode enabled.
The comparison was not made for some benchmarks with the largest inputs (indicated by dashes), as it took far too long for the sequential Eiffel programs to terminate. We remark that vector addition, dot product, and matrix-matrix multiplication have only scalar contracts; Gaussian elimination, quicksort, and matrix transposition have a combination of both scalar and range contracts (see Section \ref{sec:DbC_integration}).

There is an encouraging difference in contract-checking overhead between sequential Eiffel and \thename{}: while the former cannot maintain reasonable contract performance on larger inputs (the average slowdown for ``full'' across benchmarks with input size $10^6$, for example, is $7.19$), \thename{} has for the most part little-to-no overhead. Disabling invariant-checking leads to improvements for sequential Eiffel (which, unlike \thename{}, relies on invariant-equipped library classes), but the average slowdown is still significant (now $4.03$, for input size $10^6$). Across these benchmarks, postcondition checking adds little overhead to sequential Eiffel above checking preconditions only (which has an average slowdown of $3.98$ for input size $10^6$). \thename{} performs consistently in all modes of the experiment, with slowdown close to $1$ across the first three benchmarks. The other three benchmarks perform similarly for precondition checking, but as they include more elaborate postconditions (e.g.~``the vector is sorted''), checking both pre- and postconditions can lead to a small slowdown on large data ($1.14$ in the worst case for this experiment). Overall, the results lend support to our claim that \thename{} contracts can be monitored at runtime without diminishing the performance of the program, even with very large amounts of data. Unlike sequential Eiffel programs, contract checking need not be limited to periods of debugging.

\newcommand{\mc}[1]{\multicolumn{2}{c|}{#1}}
\newcommand{\mr}[1]{\multirow{3}{*}{#1}}

\begin{table*}[htb]
\caption{Contract checking overhead comparison}
\label{tab:contract_overhead_comparison}
\centering
{\scriptsize
\setlength{\tabcolsep}{4pt}
\begin{tabular}{ll|rr|rr|rr|rr|rr|rr}
 \mc{problem}   & \mc{$10^3$} & \mc{$10^4$} & \mc{$10^5$} & \mc{$10^6$} & \mc{$10^7$} & \multicolumn{2}{c}{$10^8$} \\
 &  & Eiffel & \thename & Eiffel & \thename & Eiffel & \thename & Eiffel & \thename & Eiffel & \thename & Eiffel & \thename  \\ \hline
\mr{Vector Addition}  
	& pre         & 1.00 & 0.92 & 1.42 & 0.96 & 3.50 & 0.96 & 3.92 & 0.95 & 3.98 & 1.02 & 4.12 & 1.06 \\
	& pre \& post & 1.00 & 0.92 & 1.42 & 0.96 & 3.66 & 0.96 & 3.93 & 0.95 & 3.98 & 1.02 & 4.29 & 1.06 \\
	& full        & 1.00 & 0.92 & 2.86 & 0.96 & 7.00 & 0.96 & 7.81 & 0.95 & 7.82 & 1.02 & 7.97 & 1.06 \\  \hline
\mr{Dot Product}
    & pre           & 1.00 & 1.02 & 1.25 & 0.99 & 4.00 & 0.97 & 3.95 & 1.01 & 4.00 & 1.10 & 4.01 & 0.95  \\
    & pre \& post   & 1.00 & 1.02 & 1.25 & 0.99 & 4.00 & 0.97 & 3.95 & 1.01 & 4.15 & 1.10 & 4.10 & 0.98  \\
    & full          & 1.00 & 1.02 & 1.88 & 0.99 & 7.25 & 0.97 & 7.33 & 1.01 & 7.46 & 1.10 & 7.48 & 0.98  \\ \hline
\mr{Matrix-Matrix Multiplication}
	& pre         & 4.00 & 1.05 & 4.47 & 1.01 & 4.55 & 0.99 & 4.54 & 0.99 &  - &  &   - &  \\
	& pre \& post & 4.00 & 1.05 & 4.47 & 1.01 & 4.59 & 0.99 & 4.57 & 0.99 &  - &  &   - &  \\
	& full        & 5.00 & 1.05 & 6.73 & 1.01 & 6.79 & 1.01 & 6.76 & 0.99 &  - &  &   - &  \\ \hline
\mr {Gaussian Elimination}  
	& pre         & 2.22 & 0.99 & 4.50 & 0.97 & 4.70 & 1.01 & 4.71 & 1.01 &    - &  &   - &   \\
	& pre \& post & 2.77 & 0.99 & 4.50 & 0.97 & 4.70 & 1.04 & 4.73 & 1.09 &    - &  &   - &   \\
	& full        & 4.44 & 0.99 & 6.67 & 0.97 & 6.96 & 1.04 & 6.96 & 1.09 &    - &  &   - &   \\ \hline
\mr{Quicksort}
	& pre         & 2.14 & 1.02 & 2.26 & 1.05 & 2.64 & 1.00 & 3.03 & 1.01 & 3.03 & 1.02 &  - &   \\
	& pre \& post & 2.28 & 1.02 & 2.27 & 1.05 & 2.70 & 1.02 & 3.02 & 1.07 & 3.04 & 1.08 &  - &   \\
	& full        & 3.64 & 1.02 & 4.14 & 1.05 & 5.07 & 1.02 & 6.38 & 1.07 & 6.49 & 1.09 &  - &   \\ \hline	
\mr{Matrix Transposition}
	& pre         & 2.00 & 1.05 & 2.06 & 1.01 & 2.40 & 1.02 & 3.71 & 1.01 &  3.86 & 1.02 &   4.02 & 1.01 \\
	& pre \& post & 2.00 & 1.05 & 2.06 & 1.01 & 2.40 & 1.03 & 3.96 & 1.11 &  4.05 & 1.12 &   4.27 & 1.14 \\
	& full        & 4.15 & 1.03 & 5.60 & 1.01 & 6.10 & 1.03 & 7.88 & 1.10 &  8.12 & 1.12 &   10.44 & 1.13 \\ \hline 	
\end{tabular}
}
\end{table*}

\section{Related Work}
\label{sec:related}

There is a vast and varied literature on general-purpose computing with GPUs. We review a selection of it, focusing on work that particularly relates to the overarching themes of \thename: the \emph{generation} of low-level GPU kernels from higher-level programming abstractions, and the \emph{correctness} of the kernels to be executed.

\subsection{GPU Programming and Code Generation}

At the C++ level of abstraction, there are a number of algorithmic skeleton and template frameworks that attempt to hide the orchestration and synchronization of parallel computation. Rather than code it directly, programmers express the computation in terms of some well-known patterns (e.g.~map, scan, reduce) that capture the parallel activities implicitly. SkePU~\cite{EnmyrenK12a}, Muesli~\cite{ErnstingK12a}, and SkelCL~\cite{SteuwerG13a} were the first algorithmic skeleton frameworks to target the deployment of fine-grained data-parallel skeletons to GPUs. While they do not support skeleton nesting for GPUs, they do provide the programmer with parallel container types (e.g.~vectors, matrices) that simplify memory management by handling data transfers automatically. Arbitrary skeleton nesting is provided in FastFlow~\cite{GoliG13a} (resp.~Marrow~\cite{MarquesPAM13a}) for pipeline and farm (resp.~pipeline, stream, loop), but concurrency and synchronization issues are exposed to the programmer. NVIDIA's C++ template library Thrust~\cite{NVIDIA-Thrust}, in contrast, provides a collection of data-parallel primitives (e.g.~scan, sort, reduce) that can be composed to implement complex algorithms on the GPU. While similar in spirit to \thename{}, Thrust lacks a number of its abstractions and container types; data can only be modeled by vectors, for example.

Higher-level programming languages benefit from a number of CUDA and OpenCL bindings (e.g.~Java~\cite{YanGS09a}, Python~\cite{KlocknerPLCIF12a}), making it possible for their runtimes to interact. These bindings typically stay as close to the original models as possible. While this allows for the full flexibility and control of CUDA and OpenCL to be integrated, several of the existing challenges are also inherited, along with the addition of some new ones; Java programmers, for example, must manually translate complex object graphs into primitive arrays for use in kernels. Rootbeer~\cite{Pratt-SzeligaFW12a}, implemented on top of CUDA, attempts to alleviate such difficulties by automatically serializing objects and generating kernels from Java code. Programmers, however, must still essentially work in terms of threads---expressed as special kernel classes---and are responsible for instantiating and passing them on to the Rootbeer system for execution on the GPU.

There are several dedicated languages and compilers for GPU programming. Lime~\cite{DubachCRBF12a} is a Java-compatible language equipped with high-level programming constructs for task, data, and pipeline parallelism. The language allows programmers to code in a style that separates computation and communication, and does not force them to explicitly partition the parts of the program for the CPU and the parts for the GPU. CLOP~\cite{Makarov-Hauswirth15a} is an embedding of OpenCL in the D language, which uses the standard facilities of D to generate kernels at compile-time. Programmers can use D variables directly in embedded code, and special constructs for specifying global synchronization patterns. The CLOP compiler then generates the appropriate boilerplate code for handling data transfers, and uses the patterns to produce efficient kernels for parallel computations. Other languages are more domain-specific than Lime and CLOP. StreamIt~\cite{ThiesKA02a}, for example, provides high-level abstractions for stream processing, and can be compiled to CUDA code via streaming-specific optimizations~\cite{HormatiSWMM11a}. A more recent example is VOBLA~\cite{BeaugnonKHBTAL14a}, a domain-specific language (DSL) for programming linear algebra libraries, restricting what the programmer can write, but generating highly optimized OpenCL code for the domain it supports. Finally, Delite~\cite{SujeethBLRCOO14a} is a compiler framework for developing embedded DSLs themselves, providing common components (e.g.~parallel patterns, optimizations, code generators) that can be re-used across DSL implementations, and support for compiling these DSLs to both CUDA and OpenCL.

A key distinction of \thename{} is the fact that GPGPU is offered to the programmer without forcing them to switch to a dedicated language in the first place: both the high-level API and the CUDA binding are made available through a library, and without need for a special-purpose compiler. Firepile~\cite{NystromWD11a} is a related library-oriented approach for Scala, in which OpenCL kernels are generated using code trees constructed from function values at runtime. Firepile supports objects, higher-order functions, and virtual methods in kernel functions, but does not support programming at the same level of abstraction as \thename{}: barriers and the GPU grid, for example, are exposed to developers.

\subsection{Correctness of GPU Kernels}

To our knowledge, \thename{} is the first GPU programming approach to integrate the specification and runtime monitoring of functional properties directly at the level of an API. Other work addressing the correctness of GPU programs has tended to focus on analyzing and verifying kernels themselves, usually with respect to concurrency faults (e.g.~data races, barrier divergence).

PUG~\cite{LiG10a} and GPUVerify~\cite{Betts_et-al15a} are examples of static analysis tools for GPU kernels. The former logically encodes program executions and uses an SMT solver to verify the absence of faults such as data races, incorrectly synchronized barriers, and assertion violations. The latter tool verifies race- and divergence-freedom using a technique, encoded in Boogie, based on tracking reads and writes in shadow memory.

Blom et al.~\cite{BlomHM14a} present a logic for verifying both data race freedom and functional correctness of GPU kernels in OpenCL. The logic is inspired by permission-based separation logic: kernel code is annotated with assertions expressing both their intended functionality, as well as the resources they require (e.g.~write permissions for particular locations). 

Other tools seek to show the presence of data races, rather than verify their absence. GKLEE~\cite{LiLSGGR12a} and KLEE-CL~\cite{CollingbourneCK14a} are two examples, based on dynamic symbolic execution. 
\section{Conclusion}
\label{sec:conclusion}
We presented \thename: a contract-based, modular, and efficient library for GPGPU, accessible for non-experts in GPU programming. 
The techniques of deferred execution and execution plan optimization helped to keep the library performance on par with raw CUDA solutions. Unlike CUDA programs, \thename programs are concise and equipped with contracts, thereby contributing to program safety.
We also found that GPU-based contracts can largely avoid the overhead of assertion checking. In contrast to classical, sequential contracts, it is feasible to monitor them outside of periods of debugging: data size is not an issue anymore.

This work can be extended in a variety of directions. First, in the current implementation, the optimizer is tailored to linear algebra and reduction/scan problems. Global optimizations could be introduced, such as changing the order of operations. Second, as shown in~\secref{evaluation}, GPU computing is not yet fast enough on ``small'' data sets. This could be resolved by introducing a hybrid computing model, in which copies of data are maintained on both the CPU and GPU. This could allow for switching between CPU and GPU executions depending on the runtime context. Finally, to provide better support for task parallelism, \thename{} could be integrated with Eiffel's concurrency model \cite{West-Nanz-Meyer15a}.

\acks

This work has been supported in part by ERC Grant CME \#291389.

\bibliographystyle{abbrvnat}

\renewcommand{\bibfont}{\normalsize}


\bibliography{references}

%
%

\end{document}